\documentclass[12pt]{article}
\usepackage[english]{babel}
\usepackage{amsfonts}
\usepackage{amsmath}
\usepackage{amssymb}
\usepackage{mathrsfs}
\usepackage{latexsym}
\usepackage{multicol}
\usepackage{fancybox} 
\usepackage{color}
\usepackage{hyperref}
\usepackage{graphicx} 
\usepackage{caption}

\usepackage{upgreek}
\usepackage[overload]{empheq}

\def\be{\begin{equation}}
\def\ee{\end{equation}}

\def\d'{``}

\newtheorem{thm}{Theorem}[section]

\newtheorem{propn}[thm]{Proposition}

\def\be{\begin{equation}}
\def\ee{\end{equation}}
\def\bea{\begin{eqnarray}}
\def\eea{\end{eqnarray}}

\def\i'{\textrm{i}}

\def\d'{``}

\begin{document}

\title{Schwarzian derivative, Painlev\'e XXV-Ermakov equation and B\"acklund transformations} 

\author{Sandra Carillo\footnote{Dipartimento Scienze di Base e Applicate per l'Ingegneria, 
\textsc{Sapienza}  Universit\`a di Roma, Roma, Italy.}, Alexander Chichurin\footnote{Institute of Mathematics, Informatics and Landscape Architecture,
          The John Paul II Catholic University of Lublin,
            ul. Konstantynow 1H, 20-708 Lublin, Poland.}, Galina Filipuk\footnote{Institute of Mathematics,  University of Warsaw,     ul. Banacha 2,   02-097 Warsaw,  Poland.}, Federico Zullo\footnote{DICATAM, Universit\`a di Brescia, Brescia, Italy.}}

\maketitle 

\begin{abstract}
The role of Schwarzian derivative in the study of nonlinear ordinary differential equations is revisited. Solutions and invariances admitted by Painlev\'e XXV-Ermakov equation, Ermakov equation and third order linear equation in a normal form are shown to be based on solutions of the Schwarzian equation. Starting from the Riccati equation and the second order element of the Riccati chian as the simplest examples of linearizable equations, by introducing a suitable change of variables, it is shown how the Schwarzian derivative represents a key tool in the construction of solutions. Two families of  B\"acklund transformations which link the linear and nonlinear equations under investigation are obtained. Some examples with relevant applications are given and discussed.
\end{abstract}


\qquad \qquad \quad ~ Painlev\'e XXV-Ermakov equation, Ermakov equation.


\section{Introduction}\label{sec1}
The Riccati equation
\begin{equation}\label{Ric}
\frac{dv}{dz}=a_2(z)v(z)^2+a_1(z)v(z)+a_0(z)
\end{equation}
can be considered as the simplest nonlinear ordinary differential equation. It is the only first-order nonlinear ordinary differential
equation which possesses the Painlev\'e property \cite{Ince}. It has many applications in different areas: from control theory to the theory of random processes, diffusion problems, orbiting satellites and seasonal phenomena \cite{Khan}, \cite{Reid}. Also, it plays a very important role in the solution
of integrable nonlinear partial differential equations. As an example, the simplest B\"acklund transformation of the Korteweg-de Vries equation is represented by a Riccati equation \cite{Levi}.  The Riccati equation (\ref{Ric}) can be linearized: indeed by setting 
\begin{equation}
v=-\frac{y'}{a_2y}
\end{equation}
it follows that $y(z)$ solves a linear second order differential equation. The linear second order differential equations are strictly related to the Schwarzian derivative. This differential operator is invariant under linear fractional transformations and plays a fundamental role in different area of mathematics: besides the theory of linear second 
order differential equations, there are numerous applications in the theory of modular forms, hypergeometric functions, univalent functions and conformal mappings. Given any smooth enough function $f(z)$, 
its Schwarzian derivative $\{f,z\}$ is defined via 
\begin{equation}\label{Schw}
\{f,z\}:=\left(\frac{f''(z)}{f'(z)}\right)'-\frac{1}{2}\left(\frac{f''(z)}{f'(z)}\right)^2 =
\frac{f'''(z)}{f'(z)}-\frac{3}{2}\left(\frac{f''(z)}{f'(z)}\right)^2.
\end{equation}
The link between the Schwarzian derivative and the theory of linear differential equations is given by the following result 
\begin{thm} (see e.g. \cite{Hille}, Theorem 10.1.1): 
if $B(z)$ is  analytic in a simply connected domain $\Omega\subset\mathbb{C}$, then for any two linearly independent solutions $\eta_1$ and $\eta_2$ of
\begin{equation}\label{eq2}
\eta''(z)=B(z)\eta(z),
\end{equation}
their quotient $\Omega=\eta_1/\eta_2$ is locally injective and satisfies the differential equation
\begin{equation}\label{Schweq2}
\{\Omega,z\}=-2B(z).
\end{equation}
\end{thm}
The converse of this statement is also true.

Higher order linear differential equations can be associated to generalized Riccati equations: all together these equations are called the \emph{Riccati chain} \cite{Levi}. In this paper we discuss the links between the second member of the Riccati chain (corresponding to a third order linear equation) and the Schwarzian derivative: in particular the connections with the Ermakov equation and with the Painlev\'e XXV-Ermakov equation studied in \cite{CF} are investigated. We show how the Painlev\'e XXV-Ermakov equation can be linearized to a third order equation written in a normal form. Thanks to the properties of this linear equation, in Section (\ref{Wro}) we find a class of auto-B\"acklund transformations for the Ermakov equation and a particular case of the Painlev\'e XXV-Ermakov equation. The Ermakov equation appears in a variety of important physical applications, including cosmology, partial differential equations of mathematical physics, elasticity, quantum mechanics and nonlinear systems (for these and other applications see e.g. \cite{Leach} and references therein).

 In Section (\ref{Wro}) polynomial relations among solutions are also introduced. In Section (\ref{secbt}) a second class of auto-B\"acklund transformations for the Ermakov equation and the Painlev\'e XXV-Ermakov equation are given: as we will see this new class of transformations is strictly related to the properties of the Schwarzian derivative. In Section (\ref{secW}) algebraic relations among the solutions of the Schwarzian equation
\begin{equation}\label{Sch}
\{\Omega,z\}=-2B(z),
\end{equation}
and its derivative (i.e. $\Omega$ and $\Omega'$)  and the solutions of the Ermakov equation and of the  Painlev\'e XXV-Ermakov equation are derived. These relations and the use of the second class of auto-B\"acklund transformations give the possibility to write the general solution of the  Ermakov equation and a particular case of the  Painlev\'e XXV-Ermakov equation in terms of $\Omega$ and $\Omega'$. Finally, in Section (\ref{Ex}) two examples are discussed: the first one involving the Weierstrass elliptic function and the second one involving a rational function with an arbitrary number of double poles.

\section{The second member of the Riccati chain and the Painlev\'e XXV-Ermakov equation}\label{secRic}
The properties of the solutions of the Riccati equation can be generalized to higher order equations: in this case the corresponding equations constitute the so called Riccati-chain \cite{Levi}. By defining the differential operator
\begin{equation}\label{diffop}
L=\frac{d}{dz}+v(z),
\end{equation}
the $n$-th order equation of the chain is represented by
\begin{equation}\label{Rc}
L^nv(z)+\sum_{k=1}^{n}a_k(z)\left(L^{k-1}v(z)\right)+a_0(z)=0,
\end{equation}
which can be linearized to a $(n+1)$-th order differential equation via the change of variables $v=y'/y$, giving
\begin{equation}\label{linear}
\frac{d^{n+1}y}{dz^{n+1}}+\sum_{k=0}^{n}a_{k}(z)\frac{d^{k}y}{dz^{k}}=0.
\end{equation}

The second order element of the Riccati chain (\ref{Rc}) is considered in \cite{CF}. When $n=2$, equation (\ref{Rc}) reads: 
\begin{equation}
\frac{d^2v}{dz^2}+3v\frac{dv}{dz}+v^3+p(z)(v'+v^2)+q(z)v+r(z)=0,
\end{equation}
where, with the notations of \cite{CF}, we set $a_2(z)=p(z)$, $a_1(z)=q(z)$ and $a_0(z)=r(z)$. The corresponding linear equation is then given by
\begin{equation}\label{lin3}
y'''+p(z)y''+q(z)y'+r(z)=0.
\end{equation}
Like for the Riccati equation (i.e. the first member of the chain), it is possible to consider the Schwarzian derivative of the ratio of two independent solutions of the corresponding linear equation. In this case, by setting 
\begin{equation}
y_1(z)=w(z)y_2(z),
\end{equation}
where both $y_1$ and $y_2$ satisfy (\ref{lin3}), we define the function $\xi(z)$ to be the Schwarzian derivative of $w(z)$, i.e.
\begin{equation}
\xi(z)\doteq \{w(z),z\}.
\end{equation}
In \cite{CF} it is shown that $\xi(z)$ satisfies the following non-linear, second order, differential equation
\begin{equation}\label{eq1}
(12\xi(z)+b(z))\xi''=15\xi'^2-h_0\xi'-8\xi^3-h_1\xi^2-h_2\xi-h_3,
\end{equation} 
where the functions $h_i$, $i=0...3$, are determined in terms of the functions $p$, $q$, $r$ and their derivatives as
\begin{equation}
\begin{split}
& b=2(p^2-3q+3p'), \; h_1=4b,\\
&h_0=2(4pp'-3p''-9pq+2p^3-6q'+27r),\\
&h_2=2(-4pp''-12qp'+2p^2p'+5p'^2+6qp'-6p^2q+p^4+6q''+9q^2-18r'),\\
&h_3=-6p''q'-2pqp''+18rp''+6p'q''+2pp'q'-2p^2qp'-2qp'^2+6q^2p'-18p'r'+\\
&+2p^2q''+2p^3q'-6pqq'+18pqr+p^2q^2-6p^2r'-4p^3r-6qq''+3q'^2+\\
&+18qr'-4q^3-27r^2.
\end{split}
\end{equation}
The following change of dependent variable
\begin{equation}
12\xi(z)+b(z)=12y(z),
\end{equation}
together with the definition of the functions $A(z)$ and $B(z)$
\begin{equation}
\begin{split}
&A(z)=\frac{1}{4}p''+\frac{1}{2}pp'-\frac{3}{4}q'+\frac{1}{9}p^3-\frac{1}{2}pq+\frac{3}{2}r,\\
&4B(z)=p'+\frac{1}{3}p^2-q,
\end{split}
\end{equation}
allow to recast the equation (\ref{eq1}) for $\xi(z)$ in a more concise form:
\begin{equation}\label{eqm}
yy''-\frac{5}{4}y'^2+\frac{2}{3}y^3+3Ay'+4By^2-2A'y-A^2=0.
\end{equation} 
The previous equation is the Painlev\'e XXV-Ermakov equation \cite{CF}: the name is due to the fact that, as shown here, by a suitable change of variables,  (\ref{eqm}) gives both  the Painlev\'e XXV equation (we are following the Ince's numbering \cite{Ince}) and the Ermakov equation.  Indeed it holds the following
\begin{propn}\label{propn1} In equation (\ref{eqm}), by setting
\begin{equation}\label{g}
y(z)=\frac{g(z)}{u(z)^4}, \quad  \textrm{with} \;\; g'=2A(z)u(z)^4,
\end{equation}
the following generalized Ermakov equation for $u(z)$ is obtained:
\begin{equation}\label{Ermg}
u''=B(z)u(z)+\frac{g(z)}{6u(z)^3}.
\end{equation}
\end{propn}
In this Proposition, with the adjective \emph{generalized} referred to the Ermakov equation we are meaning that the coefficient of $u^{-3}$ depends on $z$. The Ermakov equation has a constant term instead. Also, as it can be seen from (\ref{g}), the function $g(z)$ is related to $u(z)$ and so (\ref{Ermg}) is not strictly a Ermakov equation. Clearly, in the case $A=0$, (\ref{g}) implies that $g(z)$ is constant and in this case it follows that (\ref{Ermg}) is a proper Ermakov equation.

A particular case of the previous transformation, leading to the Painlev\'e XXV equation, is given in the following Proposition.
\begin{propn}\label{propn2} In Proposition (\ref{propn1}), by setting $g(z)=2A(z)u(z)^3$ and not considering the further constraint $g'=2Au^4$, i.e. by setting
\begin{equation}\label{g1}
y(z)=\frac{2A(z)}{u(z)}
\end{equation}
in equation (\ref{eqm}), the Painlev\'e XXV equation is obtained for $u(z)$:
\begin{equation}\label{XXV}
u''=\frac{3 u'^2}{4
   u}+  \left(\frac{A' }{2 A }-\frac{3 u }{2}\right)u'-\frac{1}{4} u^3+\frac{
   A'}{2 A}u^2+
\left(4B-\frac{5 A' {}^2}{4
   A {}^2}+\frac{A'' }{A } \right)u+\frac{4}{3}A.
\end{equation}

\end{propn}
 
From another point of view, we notice that in the Proposition (\ref{propn1}) it is possible to look at the equation (\ref{Ermg}) as an equation \emph{defining} the function $g(z)$. The constraint (\ref{g}), i.e. $g'=2A(z)u(z)^4$, then gives the following quadratic equation for $u(z)$:
\begin{equation}\label{qu}
u'''u+3u''u'-\frac{A}{3}u^2-4Buu'-B'u^2=0,
\end{equation}
which gives a \emph{linear} equation in the new variable $w=u^2$:
\begin{equation}\label{lw}
w'''-4Bw'-2\left(B'+\frac{A}{3}\right)w=0.
\end{equation}
From the previous observation we get a linearization of the equation (\ref{eqm}): indeed it holds the following
\begin{propn}\label{propn3}
Let $w(z)$ be a solution of the linear equation (\ref{lw}). Then the function
\begin{equation}\label{yw}
y=\frac{3w''}{w}-\frac{3}{2}\left(\frac{w'}{w}\right)^2-6B
\end{equation}
is a solution of the Painlev\'e XXV-Ermakov equation (\ref{eqm}). 
\end{propn}
  
\begin{propn}\label{rem1}
If $A=0$, the linear equation (\ref{lw}) reduces to the equation considered also by Gambier \cite{CM}, \cite{Hone}  in relation to the Ermakov equation. Indeed, in this case, equation (\ref{lw}) possesses a first integral given by
\begin{equation}
w''w-\frac{1}{2}(w')^2-2Bw^2=2I,
\end{equation}
where $I$ is a constant. From  Propositions (\ref{propn1}) and equation (\ref{yw}) it follows that the function $u(z)$ defined by $w=u^2$ solves the Ermakov equation:
\begin{equation}\label{ermi}
u''=B(z)u+\frac{I}{u^3}.
\end{equation}
\end{propn}
In Proposition (\ref{propn2}) we have just considered the equation for $g(z)$, without the further constraint $g'=2Au^4$. If the derivative of $g$ satisfies this  last constraint too, then it is possible to get a family of solutions of equation (\ref{eqm}), corresponding also to a family of solutions of a linear equation and of the Painlev\'e XXV equation. Indeed, from the equations $g(z)=2A(z)u^3$ and $g'=2A(z)u^4$ it follows that $u(z)$ satisfies a Riccati equation:
\begin{equation}\label{Ricu}
3A u'-Au^2+A'u=0.
\end{equation}
By setting $u=v'$, the function $A$ can be expressed in terms of $v$ and its derivative as:
\begin{equation}\label{Av}
A(z)=\frac{ce^{v}}{(v')^3},
\end{equation}
where $c$ is an arbitrary constant. By Proposition (\ref{propn1}) it follows that $u=v'$ satisfies also equation (\ref{Ermg}) (a linear equation in this case). This equation fixes the value of $B(z)$ in terms of $v$ and its derivatives. Indeed one has:
\begin{equation}\label{Bv}
B(z)=\frac{v'''}{v'}-\frac{ce^v}{3(v')^4}.
\end{equation}
Finally, the function $y(z)$ satisfying the Painlev\'e XXV-Ermakov equation is given by:
\begin{equation}\label{yv}
y(z)=\frac{2ce^{v}}{(v')^4}.
\end{equation}
From the above equations we get the following 
\begin{propn}\label{propn3.1}
Suppose that the functions $A(z)$ and $B(z)$ in (\ref{eqm}) are given by the expressions (\ref{Av}) and $(\ref{Bv})$ for some function $v(z)$. Then a solution of the Painlev\'e XXV-Ermakov equation (\ref{eqm}) is given by the expression (\ref{yv}). Further, the function $u=v'$ satisfies the Painlev\'e XXV equation (\ref{XXV}), equation (\ref{qu}) and the linear equation $u''=Bu+A/3$. 
\end{propn}

\section{Wronskians and algebraic relations among\\ solutions}\label{Wro}
The third order linear equation (\ref{lw}) is in normal form \cite{Gregus}. The function $A(z)$ is usually called the Laguerre invariant  (see e.g. at \cite{Gregus}). It is known (see again \cite{Gregus}) that, if $w_1$ and $w_2$ are two solutions of (\ref{lw}), i.e.:
\begin{equation}\label{Greg1}
w_i'''-4Bw_i'-2\left(B'+\frac{A}{3}\right)w_i=0, \quad i=1,2,
\end{equation}
then their Wronskian 
\begin{equation}\label{wro}
w=w_1w'_2-w'_1w_2
\end{equation}
is a solution of 
\begin{equation}\label{Greg2}
w'''-4Bw'-2\left(B'-\frac{A}{3}\right)w=0.
\end{equation}
In the case $A=0$ the transformation (\ref{wro}) represents an auto-B\"acklund transformation for the equation (\ref{lw}). This case is interesting since, as it has been discussed in the Proposition (\ref{rem1}), it is related to the Ermakov equation. From the B\"acklund transformation (\ref{wro}) and the changes of variables $w=u^2$, $w_i=u_i^2$, $i=(1,2)$, (see the Proposition (\ref{rem1})) we get the B\"acklund transformations for the Ermakov equation. It holds the following 
\begin{propn}\label{proposu}
Suppose that $u_1$ and $u_2$ are two solutions of the Ermakov equation
\begin{equation}\label{uc}
u''=B(z)u+\frac{c}{u^3},
\end{equation}
then the function $u$ defined by
\begin{equation}\label{uu}
u^2=2u_1u_2(u_1u'_2-u_2u'_1)
\end{equation}
is a solution of the following Ermakov equation
\begin{equation}\label{uk}
u''=B(z)u+\frac{k}{u^3}.
\end{equation}
The constants $c$ and $k$ are conserved quantities for the equation (\ref{lw}) in the case $A=0$, i.e. if $u^2=w, u_1^2=w_1$ and $u_2^2=w_2$
\begin{equation}\label{cons}
\begin{split}
& w_i''w_i-\frac{1}{2}(w_i')^2-2Bw_i^2=2c, \quad i=1,2, \\
& w''w-\frac{1}{2}(w')^2-2Bw^2=2k.
\end{split}
\end{equation}
\end{propn}

\begin{propn}\label{rem2}
From equations (\ref{uc}) and (\ref{uk}) and the relation (\ref{uu}) it follows that the functions $u_1$, $u_2$ and $u$ of the Proposition (\ref{proposu}) satisfy the following polynomial equation:
\begin{equation}\label{eqrem}
\frac{u^4}{4}+c(u_1^2-u_2^2)^2-au_1^2u_2^2=0,
\end{equation} 
where the constant $a$ is related to $k$ and $c$ by $k=-a(a+4c)$.
\end{propn}
Indeed, differentiation of equation (\ref{uu}) and taking into account the equation (\ref{uu}) itself and equation (\ref{uc}), gives
\begin{equation}\label{udiff}
uu'=c\left(\frac{u_1^2}{u_2^2}-\frac{u_2^2}{u_1^2}\right)+\frac{u^2}{2}\left(\frac{u'_1}{u_1}+\frac{u'_2}{u_2}\right).
\end{equation}
For the second derivative $u''$, by differentiating the previous equation and taking into account (\ref{uu}), (\ref{uc}) and (\ref{udiff}), we get
\begin{equation}\label{u2diff}
u''=B(z)u-\frac{(u^4+4c(u_1^2-u_2^2)^2)(u^4+4c(u_1^2+u_2^2)^2)}{16u_1^4u_2^4u^3}.
\end{equation}
By comparison with Proposition (\ref{proposu}) the ratio on the right hand side of the previous equation must be equal to $k/u^3$, i.e.,
\begin{equation}\label{polu}
\frac{(u^4+4c(u_1^2-u_2^2)^2)}{4u_1^2u_2^2}\frac{(u^4+4c(u_1^2+u_2^2)^2)}{4u_1^2u_2^2}+k=0.
\end{equation}
From direct differentiation and by using the equation (\ref{udiff}) and (\ref{uu}) it is possible to check that both the factors in equation (\ref{polu}) are constants, i.e. there exist two constants $a$ and $b$ such that: 
\begin{equation}
\begin{split}
&\frac{u^4}{4}+c(u_1^2-u_2^2)^2=au_1^2u_2^2,\\
&\frac{u^4}{4}+c(u_1^2+u_2^2)^2=bu_1^2u_2^2.
\end{split}
\end{equation} 
By subtracting the previous equations it follows that $-4c=a-b$, whereas by multiplying them one has $ab=-k$, giving the result (\ref{eqrem}).

The Propositions (\ref{proposu}) and (\ref{rem2}) can be generalized also to the case when $u_1$ and $u_2$ solve two Ermakov equations with two different constants as coefficients of the $u^{-3}$ term. We have the following Proposition.

\begin{propn}\label{rem3}
Suppose that $u_1$ and $u_2$ are two solutions of the Ermakov equation
\begin{equation}\label{uc.1}
u_i''=B(z)u_i+\frac{c_i}{u_i^3}, \quad i=1,2.
\end{equation}
Then the function $u$ defined by
\begin{equation}\label{uu.1}
u^2=2u_1u_2(u_1u'_2-u_2u'_1)
\end{equation}
is a solution of the following Ermakov equation
\begin{equation}\label{uk.1}
u''=B(z)u+\frac{k}{u^3}.
\end{equation}
The constants $c_i$ and $k$ are conserved quantities for the equation (\ref{lw}) in the case $A=0$, i.e. if $u^2=w, u_1^2=w_1$ and $u_2^2=w_2$
\begin{equation}\label{cons.1}
\begin{split}
& w_i''w_i-\frac{1}{2}(w_i')^2-2Bw_i^2=2c_i, \quad i=1,2, \\
& w''w-\frac{1}{2}(w')^2-2Bw^2=2k.
\end{split}
\end{equation}
Further, the solutions $u$, $u_1$ and $u_2$ and the constants $c_1$, $c_2$ and $k$ are related by the following polynomial relation:
\begin{equation}
\frac{u^8}{16}+\frac{(c_1u_2^4+c_2u_1^4)}{2}u^4+(c_1u_2^4-c_2u_1^4)^2+ku_1^4u_2^4=0.
\end{equation}
\end{propn}


Indeed, let us look again at the B\"acklund transformation (\ref{wro}) and the Proposition (\ref{rem1}). From the equations
\begin{equation}\label{consi}
w_i''w_i-\frac{1}{2}(w_i')^2-2Bw_i^2=2c_i, \quad i=1,2,
\end{equation}
and the change of variables $w_i=u_i^2$, $i=(1,2)$ we get that the functions $u_{i}$, $i=1,2$, solve the  Ermakov equations
\begin{equation}\label{uc12}
u_i''=B(z)u_i+\frac{c_i}{u_i^3}, \quad i=1,2
\end{equation}
whereas the function $u$, defined by
\begin{equation}\label{uu12}
u^2=2u_1u_2(u_1u'_2-u_2u'_1)
\end{equation}
solves the equation 
\begin{equation}\label{uk1}
u''=B(z)u+\frac{k}{u^3}
\end{equation}
for a suitable value of $k$. This value of $k$ is such that, if $w=u^2$, then 
\begin{equation}
w''w-\frac{1}{2}(w')^2-2Bw^2=2k.
\end{equation}
Differentiating equation (\ref{uu12}) and by taking into account the equation (\ref{uu12}) itself and equations (\ref{uc12}), we get:
\begin{equation}\label{udiff1}
uu'=\left(c_2\frac{u_1^2}{u_2^2}-c_1\frac{u_2^2}{u_1^2}\right)+\frac{u^2}{2}\left(\frac{u'_1}{u_1}+\frac{u'_2}{u_2}\right).
\end{equation}
For the second derivative $u''$, by differentiating the previous equation and taking into account (\ref{uu12}), (\ref{uk1}) and (\ref{udiff1}), we get:
\begin{equation}\label{u2diff1}
u''=B(z)u-\left(\frac{u^8+8(c_1u_2^4+c_2u_1^4)u^4+16(c_1u_2^4-c_2u_1^4)^2}{16u_1^4u_2^4u^3}\right).
\end{equation}
From equation (\ref{uk1}) it follows that the ratio on the right hand side of the previous equation must be equal to $k/u^3$, i.e.
\begin{equation}
\frac{u^8}{16}+\frac{(c_1u_2^4+c_2u_1^4)}{2}u^4+(c_1u_2^4-c_2u_1^4)^2+ku_1^4u_2^4=0.
\end{equation}

Propositions (\ref{rem2}) and (\ref{rem3}), in particular the polynomial relations among solutions, can be extended to the solutions of the third order equation (\ref{lw}) and to the solutions of the Painlev\'e XXV-Ermakov equation  (\ref{eqm})  when the function $A(z)$ is equal to zero. We have indeed
\begin{propn}\label{propos3}
Suppose that $w_1$ and $w_2$ are two solutions of the linear equation 
\begin{equation}\label{wc}
w'''-4Bw'-2B'w=0,
\end{equation}
then the function $w$ defined by the Wronskian
\begin{equation}\label{wu}
w=w_1w'_2-w_2w'_1
\end{equation}
is a solution of the same equation (\ref{wc}). There is a polynomial relation among $w$, $w_1$ and $w_2$:
\begin{equation}\label{polw}
\frac{w^4}{16}+\frac{(c_1w_2^2+c_2w_1^2)}{2}w^2+(c_2w_1^2-c_1w_2^2)^2+kw_1^2w_2^2=0.
\end{equation}
The constants $c_1$, $c_2$ and $k$ are conserved quantities for the equation (\ref{wc}), i.e.
\begin{equation}\label{cons1}
\begin{split}
& w_i''w_i-\frac{1}{2}(w_i')^2-2Bw_i^2=2c_i, \quad i=1,2, \\
& w''w-\frac{1}{2}(w')^2-2Bw^2=2k.
\end{split}
\end{equation}
If the two constants $c_1$ and $c_2$ are equal, $c_1=c_2=c$, then the polynomial equation (\ref{polw}) reduces to the quadratic equation
\begin{equation}\label{pol2w}
\frac{w^2}{4}+c(w_1-w_2)^2-aw_1w_2=0,
\end{equation}
wherein the constants $a$, $k$ and $c$ are related via $k=-a(a+4c)$.
\end{propn}
The previous Proposition comes directly from equations (\ref{Greg1})-(\ref{Greg2}) and Propositions (\ref{rem2}) and (\ref{rem3}) by substituting $u_i^2=w_i$, $i=1,2$, and $u^2=w$.

Now, let us consider the equation (\ref{eqm}) in the case $A=0$. From the Proposition (\ref{propn3}) and the Proposition (\ref{rem1}) we have that, if the functions $w_1$, $w_2$ and $w$ satisfy equations (\ref{cons}), then the functions $y_1$, $y_2$ and $y$ defined by the relations $yw^2=6k$, $y_1w_1^2=6c_1$ and $y_2w_2^2=6c_2$ satisfy equation (\ref{eqm}) in the case $A=0$. Indeed, let us consider the equation (\ref{lw}) in the case $A=0$, like in the Proposition (\ref{rem1}):
\begin{equation}\label{lw1}
w'''-4Bw'-2B'w=0.
\end{equation}
If $w$ is a solution of (\ref{lw1}), then the function defined by the transformation (\ref{yw}), i.e.,
\begin{equation}\label{yw1}
y=\frac{3w''}{w}-\frac{3}{2}\left(\frac{w'}{w}\right)^2-6B
\end{equation}
is a solution of the Painlev\'e XXV-Ermakov equation (\ref{eqm}) for $A=0$. Equation (\ref{lw1}) has a first integral, i.e. 
\begin{equation}\label{fi}
w''w-\frac{1}{2}(w')^2-2Bw^2=2c,
\end{equation}
where $c$ is a constant. By comparing equations (\ref{yw1}) and (\ref{fi}), we see that equation (\ref{yw1}) can be rewritten as
\begin{equation}\label{yw2}
y=\frac{6c}{w^2}.
\end{equation}
Now, by considering the maps $yw^2=6k$, $y_1w_1^2=6c_1$ and $y_2w_2^2=6c_2$ and expressing the relations (\ref{wu}) and (\ref{polw}) in terms of the new variables $y_1$, $y_2$ and $y$ we get the following
\begin{propn}\label{proposy}
Suppose that $y_1$ and $y_2$ are two solutions of the equation 
\begin{equation}\label{eqm1}
yy''-\frac{5}{4}y'^2+\frac{2}{3}y^3+4By^2=0,
\end{equation} 
then the function $y$ defined by the following equation
\begin{equation}\label{wy}
y=\frac{2k(y_1y_2)^3}{3c_1c_2(y_1y'_2-y_2y'_1)^2}
\end{equation}
is a solution of the same equation (\ref{eqm1}). The constants $c_1$, $c_2$ and $k$ are such that the functions $w_1$, $w_2$ and $k$ defined by $yw^2=6k$, $y_1w_1^2=6c_1$ and $y_2w_2^2=6c_2$ satisfy equations (\ref{cons1}). There is a polynomial relation among $y$, $y_1$ and $y_2$, given by:
\begin{equation}\label{poly}
16c_1c_2(c_1c_2(y_1 - y_2)^2 + ky_1y_2)y^2 + 8c_1c_2ky_1y_2(y_1+ y_2)y + k^2y_1^2y_2^2=0.
\end{equation}
\end{propn}

\section{Further B\"acklund transformations}\label{secbt}
In the previous work \cite{CZ1} (see also \cite{CZ2}) another family of B\"acklund transformations for equation (\ref{ermi}) has been presented. These transformations depend on a certain function $f(z)$ that solve a suitable functional equation that is closely related with the properties of the Schwarzian derivative. For completeness we report the result given in \cite{CZ1}:
\begin{propn}\label{ErmBT}
Suppose that $u_0$ is a solution of the equation (\ref{ermi}). Define the function $\displaystyle{\Omega(z)=\frac{\eta_0}{\eta_1}}$, where $\eta_0$ and $\eta_1$ are two independent solutions of the linear differential equation $\eta''=B\eta$. Then, if it is possible to find a function $f(z)$ such that the following equation
\begin{equation}\label{flt}
\Omega(f(z))=\frac{a\Omega(z)+b}{c\Omega(z)+d}, \qquad ad-bc\neq 0,
\end{equation}
holds for some set of constants $a,b,c,d$, then the map
\begin{equation}\label{utr}
u_1(z)^2=\frac{u_0(f(z))^2}{f'(z)}
\end{equation}
is a B\"acklund transformation for the equation (\ref{ermi}). 
\end{propn}
Indeed, let us assume that $u_0(z)$ is a solution of equation (\ref{ermi}) and hence a solution of the following equation
\begin{equation}\label{eqmfu}
\begin{split}
&\frac{d^2u(\tilde{z})}{d\tilde{z}^2}=B(\tilde{z})u(\tilde{z})+\frac{I}{u(\tilde{z})^3},\\
& \tilde{z}=f(z).
\end{split}
\end{equation} 
Differentiating equation (\ref{utr}) with respect to $z$ we find (as in equation (\ref{eqmfu}) $\tilde{z}=f(z)$)
\begin{equation}\label{1der}
u_1\frac{du_1}{dz}=u_0(\tilde{z})\frac{du_0}{d\tilde{z}}-\frac{1}{2}\frac{\frac{d^2f}{dz^2}}{(\frac{df}{dz})^2}u_0(\tilde{z})^2.
\end{equation}
The second derivative, by using equation (\ref{1der}) for $u'_1$, equation (\ref{utr}) for $u_0(\tilde{z})$ and equation (\ref{eqmfu}) for $\frac{d^2u_0(\tilde{z})}{d\tilde{z}^2}$, can be written as:
\begin{equation}\label{2der}
\frac{d^2u_1}{dz^2}=\left(\frac{df}{dz}\right)^2B(\tilde{z})u_1(z)+\frac{I}{u_1(z)^3}-\frac{1}{2}\{f(z),z\}u_1(z).
\end{equation}
It follows that $u_1(z)$ is a solution of equation (\ref{ermi}) if
\begin{equation}\label{eqdop}
\left(\left(\frac{df}{dz}\right)^2B(\tilde{z})-B(z)-\frac{1}{2}\{f(z),z\}\right)u_1(z)=0.
\end{equation}
The previous equation is an equation for $B(z)$ and can be rewritten as
\begin{equation}\label{BB}
(-2B(z))=(-2B(f(z)))\left(\frac{df(z)}{dz}\right)^2+\{f(z),z\}.
\end{equation}
We remember the transformation law of the Schwarzian derivative under composition of functions. If the function $\Omega$ depends on $z$ through the function $\psi(z)$, i.e.  $\Omega(z)=\Omega(\psi(z))$, then the Schwarzian derivative via composition of functions behaves as (see e.g. \cite{Ford})
\begin{equation}\label{Schwarzcomp}
\{\Omega(\psi(z)),z\}=\{\Omega(\psi),\psi\}\left(\frac{d\psi(z)}{dz}\right)^2+\{\psi(z),z\}.
\end{equation} 
Notably, the latter reminds equation (\ref{BB})  when the function $f$ is identified with the function $\psi$ and the function $-2B$ with the Schwarzian derivative $\{\Omega,z\}$. So we identify $-2B(z)$ with a suitable Schwarzian derivative by setting
\begin{equation}\label{position}
-2B(z):=\{\Omega(z),z\}.
\end{equation}
In terms of $\Omega$, equation (\ref{BB}) then becomes
\begin{equation}
\{\Omega(z),z\}=\{\Omega(f(z)),z\}.
\end{equation}
Further, the previous implies that $\Omega(f(z))$ and $\Omega(z)$ are related by a fractional linear transformation, i.e. there are four constants $a,b,c$ and $d$ such that
\begin{equation}
\Omega(f(z))=\frac{a\Omega(z)+b}{c\Omega(z)+d}, \qquad ad-bc\neq 0.
\end{equation}
The function $\Omega(z)$ is defined by equation (\ref{position}). Equations (\ref{eq2}) and (\ref{Schweq2}) however imply that it can be also expressed by the ratio between two independent solutions of the equation $\eta''=B\eta$. Proposition (\ref{ErmBT}) then follows.

Due to the relations between the solutions of equation (\ref{eqm1}) and those of equation (\ref{ermi}), i.e.  $y=6I/u^4$ (see Proposition \ref{propn1}), from Proposition (\ref{ErmBT}) one has immediately the following 
\begin{propn}\label{meBT}
Suppose that $y_0$ is a solution of the equation (\ref{eqm1}). Define the function $\displaystyle{\Omega(z)=\frac{\eta_0}{\eta_1}}$, where $\eta_0$ and $\eta_1$ are two independent solutions of the linear differential equation $\eta''=B\eta$. Then, if it is possible to find a function $f(z)$ such that the following equation
\begin{equation}
\Omega(f(z))=\frac{a\Omega(z)+b}{c\Omega(z)+d}, \qquad ad-bc\neq 0,
\end{equation}
holds for some set of constants $a,b,c,d$, then the map
\begin{equation}
y_1(z)=y_0(f(z))f'^2
\end{equation}
is a B\"acklund transformation admitted by equation (\ref{eqm1}). 
\end{propn}

Actually, this family of B\"acklund transformations can be extended to the more general equation (\ref{eqm}) as well by a further constraint on the function $A(z)$, again expressed in terms of a functional equation. We give the following statement and a proof.
\begin{propn}\label{meBTA}
Suppose that $y_0$ is a solution of the equation (\ref{eqm}). Define the function $\displaystyle{\Omega(z)=\frac{\eta_0}{\eta_1}}$, where $\eta_0$ and $\eta_1$ are two independent solutions of the linear differential equation $\eta''=B\eta$. Then, if it is possible to find a function $f(z)$ such that the following equation
\begin{equation}\label{Omegaeq}
\Omega(f(z))=\frac{a\Omega(z)+b}{c\Omega(z)+d}, \qquad ad-bc\neq 0,
\end{equation}
holds for some set of constants $a,b,c,d$, 
and $A(z)$ satisfies 
\begin{equation}\label{Aeq}
A(f(z))\left(\frac{df(z)}{dz}\right)^3=A(z),
\end{equation}
then the map
\begin{equation}\label{tras}
y_1(z)=y_0(f(z))f'^2
\end{equation}
is a B\"acklund transformation for the equation (\ref{eqm}). 
\end{propn}
To prove this statement, let us assume that the map (\ref{tras}) defines a new solution of equation (\ref{eqm}). Inserting (\ref{tras}) in (\ref{eqm}) and using the following equation for $y_0(f(z))$
\begin{equation}\label{eqmf}
\begin{split}
&y(\tilde{z})\frac{d^2y(\tilde{z})}{d\tilde{z}^2}-\frac{5}{4}\left(\frac{dy(\tilde{z})}{d\tilde{z}}\right)^2+3A(\tilde{z})\frac{dy(\tilde{z})}{d\tilde{z}}+\frac{2}{3}y(\tilde{z})^3+4B(\tilde{z})y(\tilde{z})^2-2\frac{dA(\tilde{z})}{d\tilde{z}}y(\tilde{z})-A(\tilde{z})^2=0,\\
& \tilde{z}=f(z),
\end{split}
\end{equation} 
we find that $y_1(z)$ is again a solution of equation (\ref{eqm}) provided  $A(z)$ and $B(z)$ satisfy the following relations:
\begin{equation}\label{two}
\{f(z),z\}+2B(z)-2B(f(z))\left(\frac{df(z)}{dz}\right)^2=0, \quad  A(f(z))\left(\frac{df(z)}{dz}\right)^3-A(z)=0.
\end{equation}
The equation for $A(z)$ is exactly (\ref{Aeq}). The equation for $B(z)$ is exactly the equation given in (\ref{eqdop}): again, with the same line of reasoning  after equation (\ref{eqdop}) we conclude that for a suitable $\Omega(z)$ 
\begin{equation}\label{position1}
-2B(z)=\{\Omega(z),z\},
\end{equation}
and if
\begin{equation}
\Omega(f(z))=\frac{a\Omega(z)+b}{c\Omega(z)+d}, \qquad ad-bc\neq 0,
\end{equation}
Proposition (\ref{meBTA}) follows.

\section{Linearization of the Wronskian equation}\label{secW}
The previous sections shows a deep relationship between the Schwarzian equation (\ref{Schweq2}), the linear equation (\ref{wc}) and the Ermakov equation (\ref{ermi}). These connections are made more explicit in this section.  It is known (see e.g. \cite{Gregus}) that a solution of equation (\ref{wc}) 
\begin{equation}\label{wc.1}
w'''-4Bw'-2B'w=0
\end{equation}
can be written as the product of two independent solutions of $\eta''=B\eta$. If $\eta_1$ and $\eta_2$ are two such solutions, for any choice of the constant $(a,b,c,d)$ one has that the function
\begin{equation}\label{gsw}
w=(a\eta_1+b\eta_2)(c\eta_1+d\eta_2)
\end{equation}
solves equation (\ref{wc.1}). If $ad-bc\neq 0$, then the previous is the general solution of equation (\ref{wc.1}) (conversely, if $ad-bc=0$ the expression on the right hand side of (\ref{gsw}) reduces to a square of a single function). We remember that equation (\ref{wc.1}) can be integrated once, giving
\begin{equation}
ww''-\frac{1}{2}(w')^2-2Bw^2=2c,
\end{equation}
where $c$ is an integration constant. By inserting the general solution (\ref{gsw}) in (\ref{wc.1}), we find that the integration constant $c$ is explicitly given by
\begin{equation}
c=-\frac{1}{2}W^2(ad-bc)^2,
\end{equation} 
where $W$ is the Wronskian of $\eta_1$ and $\eta_2$, i.e. $W=\eta_1\eta'_2-\eta'_1\eta_2$.

On the other hand, (\ref{eq2}) and (\ref{Schweq2}) imply that if $\eta_1$ and $\eta_2$ are two linearly independent solutions of the second order equation
\begin{equation}\label{eq2.1}
\eta''(z)=B(z)\eta(z),
\end{equation}
then the quotient 
\begin{equation}\label{quo}
\Omega=\frac{a\eta_1+b\eta_2}{c\eta_1+d\eta_2}, \quad ad-bc\neq 0,
\end{equation}
satisfies the Schwarzian equation
\begin{equation}\label{Schweq2.1}
\{\Omega,z\}=-2B(z).
\end{equation}
We notice that the product defined by the right hand side of equation (\ref{gsw}) is proportional to $\Omega/\Omega'$, where $\Omega$ is given by (\ref{quo}). Specifically one has:
\begin{equation}
\frac{\Omega}{\Omega'}=-(a\eta_1 + b\eta_2)(c\eta_1 + d\eta_2)/(W(ad - bc)).
\end{equation}

The previous equations, together with Proposition (\ref{rem1}) and equation (\ref{yw2}), give the following
\begin{propn}\label{propfin}
Suppose that the function  $\Omega(z)$ solves the Schwarzian equation
\begin{equation}
\{\Omega,z\}=-2B(z).
\end{equation}
Then the following statements hold:
\begin{enumerate}
\item The function $w=\Omega/\Omega'$ solves the linear equation
\begin{equation}\label{eq3.10}
w'''-4Bw'-2B'w=0.
\end{equation}
\item  The function defined by $u^2=2r \left(\Omega/\Omega'\right)$ solves the Ermakov equation
\begin{equation}\label{eq3.20}
u''=Bu-\frac{r^2}{u^3}.
\end{equation}
\item The function $y(z)=-\frac{3}{2}\left(\Omega'/\Omega\right)^2$ satisfies the equation (\ref{eqm}) for $A=0$, i.e.
\begin{equation}\label{eq3.30}
yy''-\frac{5}{4}y'^2+\frac{2}{3}y^3+4By^2=0.
\end{equation} 
\end{enumerate}
\end{propn}
In Proposition (\ref{propfin}), particular solutions of the equations (\ref{eq3.10})-(\ref{eq3.30}) depend on the function $\Omega(z)$ and its first derivative: this function is the same appearing in the B\"acklund transformations found in section (\ref{secbt}). This observation helps to find the \emph{general} solutions of equations  (\ref{eq3.10})-(\ref{eq3.30}) in terms of $\Omega(z)$ and its first derivative. Indeed, let us consider for example the Ermakov equation (\ref{eq3.20}). From Proposition (\ref{ErmBT}) we know that if $u_0(z)$ is a solution of (\ref{eq3.2}), then 
\begin{equation}\label{utr.1}
u_1(z)^2=\frac{u_0(f(z))^2}{f'(z)}
\end{equation}
is another solution. Here the function $f(z)$ is such that 
\begin{equation}\label{flt.1}
\Omega(f(z))=\frac{a\Omega(z)+b}{c\Omega(z)+d}, \qquad ad-bc\neq 0.
\end{equation}
Differentiating the previous equation we get
\begin{equation}\label{flt.2}
\Omega'(f(z))f'(z)=(ad-bc)\frac{\Omega'(z)}{(c\Omega(z)+d)^2}, \qquad ad-bc\neq 0.
\end{equation}
Setting $u_0^2(z)=2r\left(\Omega/\Omega'\right)$ in (\ref{utr.1}) it follows
\begin{equation}\label{utr.2}
u_1(z)^2=2r\frac{\Omega(f)}{\Omega'(f)f'(z)},
\end{equation}
and, with the help of equation (\ref{flt.2}), we get
\begin{equation}\label{utr.3}
u_1(z)^2=2r\frac{(a\Omega(z)+b)(c\Omega(z)+d)}{(ad-bc)\Omega'(z)}.
\end{equation}
Equation (\ref{utr.3}) represents a solution of the Ermakov equation (\ref{eq3.20}) for any choice of the constants $(a,b,c,d)$ such that $ad-bc\neq 0$ and then it is the general solution of equation (\ref{eq3.20}). The same line of reasoning can be applied to equations (\ref{eq3.10}) and (\ref{eq3.30}): the results are summarized in the following
\begin{propn}\label{propfin1}
Suppose that the function  $\Omega(z)$ solves the Schwarzian equation
\begin{equation}
\{\Omega,z\}=-2B(z)
\end{equation}
and let the arbitrary constants $(a,b,c,d)$ be such that $ad-bc \neq 0$. Then 
\begin{enumerate}
\item The general solution of the linear equation
\begin{equation}\label{eq3.1}
w'''-4Bw'-2B'w=0
\end{equation}
is given by
\begin{equation}
w(z)=\frac{(a\Omega(z)+b)(c\Omega(z)+d)}{(ad-bc)\Omega'(z)}.
\end{equation}
\item  The general solution of the Ermakov equation
\begin{equation}\label{eq3.2}
u''=Bu-\frac{r^2}{u^3}
\end{equation}
is given by
\begin{equation}
u(z)^2=2r\frac{(a\Omega(z)+b)(c\Omega(z)+d)}{(ad-bc)\Omega'(z)}.
\end{equation}
\item The general solution of equation (\ref{eqm}) for $A=0$, i.e.
\begin{equation}\label{eq3.3}
yy''-\frac{5}{4}y'^2+\frac{2}{3}y^3+4By^2=0
\end{equation} 
is given by
\begin{equation}
y(z)=-\frac{3}{2}\left(\frac{(ad-bc)\Omega'(z)}{(a\Omega(z)+b)(c\Omega(z)+d)}\right)^2.
\end{equation}
\end{enumerate}
\end{propn}

The results given in this section will be useful in the examples of the next section.

\section{Examples and applications}\label{Ex}
This Section collects some examples of interest in applications.
\subsection{An example involving the Weierstrass $\wp$ function}
Let us consider the duplication formula for the Weierstrass elliptic function $\wp(z)$ (see e.g. \cite{NIST}, formula 23.10.7): 
\begin{equation}
\wp(2z)=\frac{1}{4}\left(\frac{\wp''(z)}{\wp'(z)}\right)^2-2\wp(z).
\end{equation}
From the differential equation satisfied by $\wp(z)$, i.e.
\begin{equation}\label{dwp}
(\wp')^2=4\wp^3-g_2\wp-g_3,
\end{equation}
where $g_2$ and $g_3$ are the invariants, and the differential consequences of (\ref{dwp}), i.e.
\begin{equation}
\wp''=6\wp^2-\frac{1}{2}g_2, \quad \wp'''=12\wp\wp',
\end{equation}
we get
\begin{equation}\label{wdouble}
\frac{\wp'''}{\wp'}-\frac{3}{2}\left(\frac{\wp''}{\wp'}\right)^2=\{\wp(z),z\}=-6\wp(2z).
\end{equation}
The previous equation and Proposition \ref{propfin1} suggests to set $B(z)=3\wp(2z)$. So we consider the linear third order equation:
\begin{equation}\label{wbw}
w'''-12\wp(2z)w'-12\wp'(2z)w=0.
\end{equation} 
From equation (\ref{wdouble}) we see that 
\begin{equation}\label{Owp}
\Omega(z)=\wp(z),
\end{equation}
whereas Proposition \ref{propfin1} tells us that the general solution of (\ref{wbw}) is given by
\begin{equation}
w(z)=\frac{(a\wp(z)+b)(c\wp(z)+d)}{(ad-bc)\wp'(z)}.
\end{equation}
If we consider the Ermakov equation (\ref{eq3.2}) in the case $B(z)=3\wp(2z)$, i.e.,
\begin{equation}\label{ermwp}
u''=3\wp(2z)u-\frac{r^2}{u^3},
\end{equation}
we get, from Proposition \ref{propfin1} the general solution
\begin{equation}
u^2(z)=2r\frac{(a\wp(z)+b)(c\wp(z)+d)}{\wp'(z)(ad-bc)}.
\end{equation}
In addition, the general solution of the equation
\begin{equation}
yy''-\frac{5}{4}y'^2+\frac{2}{3}y^3+12\wp(2z)y^2=0,
\end{equation} 
is given by
\begin{equation}
y(z)=-\frac{3}{2}\left(\frac{(ad-bc)\wp'(z)}{(a\wp(z)+b)(c\wp(z)+d)}\right)^2.
\end{equation}

We finally remark that equation (\ref{wbw}) is related to a particular case of the Lam\'e equation considered by Halphen (see \cite{Halphen}, pag. 105):
\begin{equation}\label{LameH}
\eta'' - 3/4 \wp(z) \eta = 0. 
\end{equation}
Halphen showed that this equation is such that every solution is multi-valued but the ratio of any two solutions is single-valued (a \emph{fonction uniforme} according to Halphen). Further, he gives explicitly the two independent solution of (\ref{LameH}) as 
\begin{equation}
\eta_1 = \wp'(z / 2)^{-\frac{1}2},\quad \eta_2 = \wp'(z/2)^{-\frac{1}2}\wp(z / 2).
\end{equation}
The link between equation (\ref{wbw}) and (\ref{LameH}) is obtained by considering the change of variables $2z\to z$ in (\ref{wbw}), giving
\begin{equation}\label{eqqq}
w''' -3 \wp(z) w' -3/2 \wp'(z) w = 0, 
\end{equation}
and by considering the comment after equation (\ref{wc.1}) and equation (\ref{gsw}). Indeed, from equation (\ref{gsw}), we see that the general solution of (\ref{eqqq}) can be written as
\begin{equation}
w(z)=(a\eta_1(z)+b\eta_2)(c\eta_2(z)+d\eta_2), \; ad-bc \neq 0,
\end{equation} 
where $\eta_1$ and $\eta_2$ are two independent solutions of the Lam\'e equation (\ref{LameH}). Explicitly one has
\begin{equation}
w(z)=\frac{1}{\wp'(z)}(a\wp(z)+b)(c\wp(z)+d), \; ad-bc \neq 0.
\end{equation}

\subsection{An example involving rational functions}
The Ermakov equation with  a rational potential $B(z)$, like $z^{-2}$, appears in the theory of scalar field cosmologies (see e.g. \cite{CZ1}, \cite{HL}). 
It is well known \cite{Hille} that the Schwarzian derivative of a power, say $(z-\alpha)^{n+1}$, is proportional to $(z-\alpha)^{-2}$: more precisely
\begin{equation}
\{(z-a)^{n+1},z\}=-\frac{n(n+2)}{2}\frac{1}{(z-a)^2}.
\end{equation}
 Actually it is possible to generalize the previous equation to a product of such power functions. Let us define the function $P(z)$ by
\begin{equation}\label{Pdef}
P'(z)=\prod_{k=1}^N(z-a_k)^n,
\end{equation}  
where all $a_k$'s are supposed to be different from each other. It is easy to show that
\begin{equation}\label{di1}
\frac{P'''}{P'}=n(n-1)\sum_{k=1}^N\frac{1}{(z-a_k)^2}+2n^2\sum_{k=1}^N\sum_{j\neq k}^N\frac{1}{a_j-a_k}\frac{1}{z-a_k},
\end{equation}
and
\begin{equation}\label{di2}
\left(\frac{P''}{P'}\right)^2=n^2\sum_{k=1}^N\frac{1}{(z-a_k)^2}+2n^2\sum_{k=1}^N\sum_{j\neq k}^N\frac{1}{a_j-a_k}\frac{1}{z-a_k}.
\end{equation}
Equations (\ref{di1}) and (\ref{di2}) give
\begin{equation}
\{P(z),z\}=-\frac{n(n+2)}{2}\sum_{k=1}^N\frac{1}{(z-a_k)^2}-n^2\sum_{k=1}^N\sum_{j\neq k}^N\frac{1}{a_j-a_k}\frac{1}{z-a_k}.
\end{equation}
From the Proposition (\ref{propfin1}) we get the following result. Let the function $P(z)$ be defined by (\ref{Pdef}) and consider the following third order linear equation:
\begin{equation}\label{wbP}
w'''-4Bw'-2B'w=0,
\end{equation} 
where 
\begin{equation}\label{Brat}
B(z)=\frac{n(n+2)}{4}\sum_{k=1}^N\frac{1}{(z-a_k)^2}+\frac{n^2}{2}\sum_{k=1}^N\sum_{j\neq k}^N\frac{1}{a_j-a_k}\frac{1}{z-a_k}.
\end{equation}
Then, the general solution of equation (\ref{wbP}) is given by
\begin{equation}
w(z)=\frac{(aP(z)+b)(cP(z)+d)}{(ad-bc)P'(z)}=\frac{(a\int \prod_{k=1}^N(z-a_k)^ndz+b)(c \int \prod_{k=1}^N(z-a_k)^ndz+d)}{(ad-bc)\prod_{k=1}^N(z-a_k)^n}.
\end{equation}
By considering the Ermakov equation
\begin{equation}
u''=Bu-\frac{r^2}{u^3},
\end{equation}
with $B(z)$ defined by (\ref{Brat}), its general solution is defined by the equation 
\begin{equation}
u(z)^2=2r\frac{(aP(z)+b)(cP(z)+d)}{(ad-bc)P'(z)}=2r\frac{(a\int \prod_{k=1}^N(z-a_k)^n dz+b)(c\int \prod_{k=1}^N(z-a_k)^n dz+d)}{(ad-bc)\prod_{k=1}^N(z-a_k)^n}.
\end{equation}
Finally, the general solution of the following equation
\begin{equation}
yy''-\frac{5}{4}y'^2+\frac{2}{3}y^3+4By^2=0,
\end{equation} 
where $B(z)$ is defined by (\ref{Brat}), is given by
\begin{equation}
\begin{split}
y(z)&=-\frac{3}{2}\left(\frac{(ad-bc)P'(z)}{(aP(z)+b)(cP(z)+d)}\right)^2=\\
&=-\frac{3}{2}\left(\frac{(ad-bc)\prod_{k=1}^N(z-a_k)^n}{(a\int \prod_{k=1}^N(z-a_k)^n dz+b)(c\int \prod_{k=1}^N(z-a_k)^n dz+d)}\right)^2.
\end{split}
\end{equation}

\begin{center} {\bf Acknowledgments} \end{center}
SC acknowledges the support of \textsc{Sapienza} Universit\`a di Roma, GNFM-INdAM and INFN, GF acknowledges the support of National Science Center (Narodowe Centrum Nauki NCN) OPUS grant 2017/25/B/BST1/00931 (Poland), FZ acknowledges the support of Universit\`a di Brescia,  GNFM-INdAM and INFN.
\bigskip

\begin{center} {\bf Conflict of interests} \end{center}
On behalf of all authors, the corresponding author states that there is no conflict of interest. 
\bigskip 

\begin{center} {\bf Data Availability Statements} \end{center}
Data sharing not applicable to this article as no datasets were generated or analysed during the current study.

\end{document}